\documentclass[12pt]{article}
\usepackage{amsfonts}
\usepackage{amsmath}
\usepackage{amssymb}
\usepackage{color}

\def\appendix#1
{ \addtocounter{section}{1}
\setcounter{equation}{0}
 \renewcommand{\thesection}{\Alph{section}}
 \section*{
 \thesection\protect\indent \parbox[t]{16.715cm}{#1}}
 \addcontentsline{toc}{section}{Appendix \thesection\ \ \ #1}
}

\newcommand{\kM}{$\kappa$-Minkowski}
\newcommand{\kP}{$\kappa$-Poincar\'e}

\newcommand{\bt}{\beta}
\newcommand{\kx}{{\hat{\bf x}}}
\newcommand{\qd}{\right]}
\newcommand{\qs}{\left[}

\newcommand{\de}{\mathrm{d}}
\def\nn{\nonumber}
\def\be{\begin{equation}}
\def\ee{\end{equation}}
\def\bea{\begin{eqnarray}}
\def\eea{\end{eqnarray}}

\begin{document}
\begin{titlepage}
\begin{center}
\vskip .2in {\large\bf Covariant formulation of Noether's Theorem for translations on \kM\ spacetime}\\
\vskip .2in

Alessandra AGOSTINI\footnote{e-mail: alessandraagostini@libero.it}\\
\vskip .2in
{\it Cardiff School of Mathematics, Cardiff University,\\
Sengennydd Road, Cardiff, CF24 4AG, United Kingdom} \vskip .5in

\abstract The problem of finding a formulation of Noether's theorem
in noncommutative geometry is very important in order to obtain
conserved currents and charges for particles in noncommutative
spacetimes. In this paper, we formulate \emph{Noether's theorem} for
translations of \kM\ noncommutative spacetime on the basis of the
\emph{5-dimensional \kP\ covariant} differential calculus. We focus
our analysis on the simple case of free scalar theory. We obtain
\emph{five} conserved Noether currents, which give rise to
\emph{five} energy-momentum charges. By applying our result to plane
waves it follows that the energy-momentum charges satisfy a
special-relativity dispersion relation with a generalized mass given
by the fifth charge. In this paper we provide also a rigorous
derivation of the equation of motion from Hamilton's principle in
noncommutative spacetime, which is necessary for the Noether
analysis.



\end{center}
\end{titlepage}

\section{Introduction}
Noncommutative spacetime was introduced in
Ref.~\cite{Pauli,Snyder} in order to improve the singularity of
quantum field theory at short distances. Afterwards, the idea of
noncommutative structure of spacetime inspired several approaches to
Quantum Gravity\cite{garay}. In particular, Doplicher \emph{et
al.}~\cite{DFR1,DFR2} explored the possibility that Quantum Gravity
corrections can be described algebraically by replacing the
traditional (Minkowski) spacetime coordinates $x_\mu$ by Hermitian
operators $\kx_\mu$ ($\mu = 0, 1, 2, 3$) which satisfy the
nontrivial commutation relations 
\be
[\kx_\mu,\kx_\nu]=i\theta_{\mu\nu}(\kx). \label{DFR}\ee

The model of $\kappa$-deformed spacetime considered in this paper
and denoted by the name of \emph{\kM\ spacetime} is a particular
realization of Eq.~(\ref{DFR}). It is characterized by the
Lie-algebra commutation relations $$ [\kx_0, \kx_j ] = i\lambda
\kx_j,\quad [\kx_j , \kx_k] = 0,\quad j = 1, 2, 3, $$ where the
noncommutative parameter\footnote{Historically, the noncommutative
parameter $\kappa=\lambda^{-1}$ was introduced. This explains the
origin of the name ``\kM".} $\lambda\in \mathbb{R}^+$ is usually
expected to be of the order of the Planck length.

Such a model was introduced in Ref.~\cite{LNRT}-\cite{KS} and
has been widely studied both from a mathematical and a physical
perspective\cite{Kosinski:1999dw,Dimitrijevic:2003wv}.

\kM\ algebra was also proposed in the framework of the Planck scale
Physics~\cite{Maj88MajPhDmaj00,Maj00} as a natural candidate for a
quantized spacetime in the zero-curvature limit.

Recently, $\kappa$-Minkowski gained remarkable attention due to the
fact that its symmetries have been proven to be described in terms
of a well-known deformation of the Poincar\'{e} group, the \kP\ Hopf
algebra\cite{MajidRuegg}-\cite{aad03}, which has been derived by
contracting the Hopf algebra $SO_q(3, 2)$.

The conserved charges associated with the \kP\ Hopf-algebra
transformations have been characterized on the basis of various
heuristic arguments \cite{lukieAnnPhys}. In particular, the
identification of the energy-momentum charges with the generators of
the \kP\ translations has led to hypothesis that in \kM\ sapcetime
particles may be submitted to dispersion relations modified with
respect to the Einstein ones by the presence of
$\lambda$-corrections\cite{AmelinoLukNowik1,AmelinoLukNowik2}. If
the parameter $\lambda$ is identified with the Planck length, the
modification of the dispersion relations would agree with the the
results of DSR theories which predict the existence of two
observer-independent quantities: a velocity scale and a length scale
(given by the Planck length)\cite{dsr1a,dsr1b}.

Because of these implications at the level of fundamental physics,
there is a great interest in searching for a  robust
characterization of the conserved charges associated with the \kP\
symmetry transformations, especially those associated to the \kP\
translations, which would have the meaning of energy and momenta of
particles.

The first attempt in this direction seems to have appeared in
Ref.~\cite{kappanoether1}, where a Noether analysis has been applied to a
free scalar field theory in \kM\ spacetime.

This study solved the issue about the ambiguity among different but
equivalent bases of \kP\ by proving that they give rise to the same
energy-momentum charges. However, the analysis in Ref.~\cite{kappanoether1} is restricted to the class of four-dimensional
differential calculi which are non-$\kappa$-Poincar\'{e} covariant.
In light of this, it seems natural to look for a formulation of the
Noether's theorem based on a \kP\ covariant differential calculus.
Such a covariant differential calculus is proven to be uniquely
defined and coincides with the five-dimensional one constructed by
Sitarz, see Ref.~\cite{Sitarz}.

In this paper we plan to apply a generalization of Noether's theorem
to the \kM\ translations of a free scalar theory\footnote{In this
paper the analysis of Noether's theorem is restricted to the
classical field theory in order to avoid the further complications
that arise in quantum field theory, which we postpone to future
studies.}. In order to do this we introduce the notion of \emph{\kP\
covariant} translation in \kM\ spacetime based on the
five-dimensional differential calculus. By using the
five-dimensional vector derivatives we introduce also a Lagrangian
which gives rise to \kP-invariant equation of motion. By requiring
the invariance of the action of the theory under the covariant
spacetime translations we obtain \emph{five} conserved charges. It
is important to notice that the choice of a \kP\ covariant
Lagrangian (in the sense that it produces \kP\ invariant equation of
motion) and the use of a covariant differential calculus assure
\emph{step by step} the covariance of the formulation of Noether's
theorem.

By applying our results to \kM\ \emph{plane waves} we obtain a
dispersion relation for the conserved charges. It seems to be
interesting that such a dispersion relation looks like the
special-relativity (\emph{i.e.} non-Plank-deformed) dispersion
relation in which the mass is replaced by the fifth
charge\footnote{A result similar to ours was obtained,
independently, by G. Amelino-Camelia \emph{et. al} in Ref.~\cite{AmelinoCamelia:2007vj}.}. In the case of a massless theory the
fifth conserved charge is zero and the \kM\ plane waves satisfy
exactly the special-relativity dispersion relation.

In conclusion, the result that we have obtained in this paper seems
to disagree with the deformed dispersion relation conjectured on the
basis of heuristic arguments and widely used in literature so far.

\section{\kM\ Spacetime and \kP\ Hopf-algebra Symmetry}

The coordinates of the four-dimensional \kM\ spacetime satisfy the
commutation relations of Lie-algebra type \be [\kx_0,\kx_j]=i\lambda
\kx_j,\quad [\kx_j,\kx_k]=0,\quad j,k=1,2,3,\label{kM} \ee where $\kx_0$
has the meaning of time and $\kx_j$ have the meaning of space
coordinates. In the commutative limit $\lambda\to 0$, \kM\ reduces
to the commutative Minkowski spacetime.

The set of coordinates $\kx_\mu$ and the Lie-algebra relation~(\ref{kM}) define the associative algebraic structure
$\mathcal{A}_x$ of \kM. We can consider different bases for the
algebra $\mathcal{A}_x$ which are all
equivalent~(see Refs.~\cite{Agostini:2002de}, \cite{Meljanac:2007xb}). In this
paper we shall use the time-to-the-right basis $$
\hat{e}_k=\{e^{-ik\kx}e^{ik_0\kx_0}\},\;\;\;k^\mu\in R^{(1,3)}.$$
where $\kx_0$ has the meaning of time. The product associated to this
basis is
$$ \hat{e}_k \hat{e}_p=\hat{e}_{(k\oplus p)},$$ where the
non-Abelian sum is $(k\oplus p)_\mu=(k_0+p_0, k_j+e^{-\lambda
k_0}p_j)$.

Notice the following conjugation property \be
e_k^{\dag}=e_{\dot{-}k},\label{conj}\ee where $\dot{-}k_\mu$ is
called antipode and corresponds to $(-k_0,-e^{\lambda k_0}k_j)$.

We consider fields in \kM\ spacetime as elements of the algebra
$\mathcal{A}_x$ \be \Phi(\kx)=\int \de^4k\; \mu_{\lambda}(k)\,
\tilde{\phi}(k)\, \hat{e}_k,\label{Phi}\ee where $\tilde{\phi}(k)$
is the Fourier transform of the commutative limit of $\Phi(\kx)$, and
$\mu_\lambda(k)$ is an integration measure which is equal to $1$ in
the commutative limit. The expression of $\Phi(\kx)$ is the
generalization of a classical field which is usually represented as
a Fourier expansion in plane waves. Here we do not discuss the
algebraic properties of the \kM\ functions $\Phi(\kx)$ for which we
refer to Ref.~\cite{Agostini:2005mf}, where a rigorous analysis has been
done on the representations of the \kM\ functions on Hilbert spaces
and their $C*$-algebra properties.

In Ref.~\cite{MajidRuegg} it has been proven that \kM\
noncommutative spacetime is the invariant space\footnote{The
invariance of \kM\ under \kP\ Hopf-algebra means that the
commutation relations of \kM\ are left invariant under the action of
the generators of the \kP\ algebra $$ T\rhd[\kx_0,\kx_j]=i\lambda
T\rhd\kx_j$$ where $T=(P_\mu,N_j,M_j)$. See Ref.\cite{Majid_book}
for the mathematical definition of action.} of a quantum deformation
of the Poincar\'e algebra, called \kP\ algebra, originally obtained
as a contraction of the Hopf algebra $SO_q(3, 2)$, see
Ref.~\cite{LNRT}-\cite{KS}.

The \kP\ Hofp-algebra can be written in a number of different bases.
In the so called Majid-Ruegg bicrossproduct basis the generators
$(P_\mu,N_j,M_j)$ of \kP\ satisfy the following commutation
relations \bea
\qs P_{\mu},P_{\nu}\qd&=&0,\nn\\
\qs M_j,M_k\qd&=&i\varepsilon_{jkl}M_l,\;\;\;\qs {N}_j,M_k\qd
=i\varepsilon_{jkl}{N}_l,\;\;\;\qs {N}_j,{N}_k\qd
=-i\varepsilon_{jkl}M_l,\nn\\
\qs M_j,P_0\qd&=&0,\;\;\;[M_j,P_k]=i\varepsilon_{jkl}P_l,\nn\\
\qs {N}_j,P_0\qd&=&iP_j,\nn\\
\qs {N}_j,P_k\qd&=&i\qs \left( \frac{1- e^{-2\lambda
P_0}}{2\lambda}+\frac{\lambda}{2}\vec{P}^2\right)\delta_{jk}
-\lambda P_jP_k\qd, \label{kP} \eea and the following co-algebra
relations \bea \Delta(P_0)&=&P_0\otimes 1+1\otimes
P_0\;\;\;\Delta(P_j)
=P_j\otimes 1+e^{-\lambda P_0}\otimes P_j,\nn\\
\Delta(M_j)&=&M_j\otimes 1+1\otimes M_j,\nn\\
\Delta({N}_j)&=&{N}_j\otimes 1+ e^{-\lambda P_0}\otimes
{N}_j-\lambda\varepsilon_{jkl}P_k\otimes M_l ~.\label{kPcopr} \eea

The ``mass-squared" Casimir operator in the Majid-Ruegg
bicrossproduct basis takes the form \be
C_{\lambda}(P)=\frac{2}{\lambda^2}\sinh^2\left(\frac{\lambda
P_0}{2}\right)-e^{\lambda P_0}\vec{P}^2.\label{casimir} \ee

The symmetry generators of the \kP\ Hopf algebra act in the
following way on \kM\ functions: \bea P_\mu \hat{e}_k&=&k_\mu\hat{e}_k,\label{P}\\
M_j\hat{e}_k&=&i\varepsilon_{jlm}k_l\partial_m\hat{e}_k,\nn\\
N_j\hat{e}_k&=&i\left(k_j\partial_0-(\frac{1-e^{-2\lambda
k_0}}{2\lambda}+\frac{\lambda}{2}k^2)\partial_j+\lambda
k_jk_l\partial_l\right)\hat{e}_k.\nn \eea

As the reader can note, in the bicross-product basis the generators
of the $\kappa$-Lorentz algebra fulfil the commutation relations of
the un-deformed Lorentz Lie-algebra. Nevertheless, the symmetry
generators of the \kP\ Hopf algebra act in a deformed way on
products of functions.

\section{The 5D \kP\--invariant Differential Calculus}

The issue of finding differential calculi related to \kM\ spacetime
has been investigated in different
papers\cite{Madore:1995mu,papers}. Sitarz\cite{Sitarz} proved that
there are no 4D \kP--covariant differential calculi, and proposed a
5D differential calculus which is covariant under the left action of
the  \kP\ Hopf algebra \footnote{or, equivalently, under the
infinitesimal left action of \kP\ quantum group
$\mathcal{P}_{\kappa}$, see Ref.~\cite{Zacrewski}.}. Then, Gonera
\emph{et al.}\cite{Gonera} showed that the lowest dimensional
left-covariant calculus for the 4D \kM\ spacetime is uniquely
defined and coincides with the 5D calculus proposed by Sitarz.

In the 5D differential calculus the exterior derivative operator
$\de$ of a generic \kM\ element $\Phi(\kx)$ can be written in terms
of vector fields $\mathcal{D}_a(P)$ as follows: \bea
\de\Phi(\kx)&=&\de x^a\mathcal{D}_a
\Phi(\kx),\;\;\;a=0,1,2,3,4, \label{5Ddiffcalc}\\
\mathcal{D}_0(P)&=&\frac{i}{\lambda}\left[\sinh(\lambda
P_0)+\frac{\lambda^2}{2}e^{\lambda P_0}\vec{P}^2\right],\nn\\
\mathcal{D}_j(P)&=&iP_je^{\lambda P_0}\quad (j=1,2,3), \nn\\
\mathcal{D}_4(P)&=&\frac{1}{\lambda}\left[2\sinh^2\left(\frac{\lambda
P_0}{2}\right)-\frac{\lambda ^2}{2}\vec{P}^2e^{\lambda
P_0}\right],\nn\eea where $P_\mu$ are the generators of \kP\ in the
bicrossproduct basis and act  on \kM\ as in Eq.~(\ref{P}) $ P_\mu
\hat{e}_k=k_\mu \hat{e}_k$. Notice that the last component
$\mathcal{D}_4(P)$ coincides with the Casimir operator~(\ref{casimir}).

The 5D differential calculus is obtained in Ref.~\cite{Sitarz} by the
request that that the commutation relations~(\ref{kM}) remain
invariant under the action of the \kP\ generators:
$$
[dx^a, \kx^\mu]=\rho^{a\mu}_\nu\kx^\nu \quad\to\quad  T\;[dx^a,
\kx^\mu]=\rho^{a\mu}_\nu T\;\kx^\nu$$ where T denotes globally the
\kP\ generators $(P_\mu ;M_j ;N_j)$. A differential calculus in
which the commutation relations between the 1-form generators and
the \kM\ generators remain invariant under the action of symmetry
algebra (\kP\ in our case), is called ``covariant" differential
calculus.

Clearly, the \kM\ derivatives $\mathcal{D}_a$ reduce to the
commutative derivatives $\partial_\mu$ in the limit $\lambda\to 0$:
$$
\lim_{\lambda \to
0}\mathcal{D}_\mu(P)=iP_\mu=\partial_\mu,\quad\lim_{\lambda\to
0}\mathcal{D}_4(P)=0.
$$

The deformed derivatives $\mathcal{D}_a(P)$ have some nice
covariance properties. They transform in the classical way under the
\kP\ action \bea
\qs M_j,\mathcal{D}_\mu\qd&=&i\delta_{\mu k}\epsilon_{jkl}\mathcal{D}_l,\nn\\
\qs {N}_j,\mathcal{D}_\mu\qd&=&iD_\mu,\nn\\
\qs{M_j},\mathcal{D}_4\qd&=&\qs{N_j},\mathcal{D}_4\qd=0.\nn
 \eea
Their coproduct can be written as \bea
\Delta(\mathcal{D}_\bt)&=&\mathcal{D}_\bt \otimes e^{\lambda
P_0}+e^{-\lambda P_0\delta_{\bt0}}\otimes \mathcal{D}_\bt
-i\delta_{\bt0}\lambda e^{-\lambda P_0}\mathcal{D}_j \otimes
\mathcal{D}_j,\nn\\
\Delta(\mathcal{D}_4)&=&\mathcal{D}_4\otimes e^{\lambda
P_0}+e^{-\lambda P_0}\otimes
\mathcal{D}_4+\frac{1}{\lambda}e^{-\lambda P_0}(e^{\lambda
P_0}-1)\otimes(e^{\lambda P_0}-1)\nn\\
&&+\lambda e^{-\lambda P_0}\mathcal{D}_j\otimes
\mathcal{D}_j.\label{copr_D_mu} \eea

The one-form generators $\de x^a=(dx^\mu, dx^4)$ satisfy the
commutation relations (see Ref.~\cite{Sitarz})
\bea [\de x^{\mu},\kx^\nu]&=&i\lambda(g^{0\mu}\de x^\nu-g^{\mu\nu}\de x^0+ig^{\mu\nu}\de x^4), \nn\\
{}[\de x^4,\kx^\mu]&=&\lambda \de x^\mu.\label{diff_comm_rel}\eea The
fifth one-form generator is here denoted by $\de x^4$, but this is
of course only a formal notation since there is no fifth \kM\
coordinate $\kx^4$.

The one-forms $\de x^a$ have the following hermitian properties:
$$(\de x^{\mu})^*=\de x^\mu,\quad (\de x^4)^*=-\de x^4.$$

The external algebra takes the standard form $$ \de x^a\wedge\de
x^b=-\de x^b\wedge\de x^a\quad (a,b=0,1,2,3,4),$$ and $$ \de(\de
x^\mu)=0,\quad\de(\de x^4)=-\frac{2i}{\lambda}\de(\de x^\mu)\wedge
\de(\de x_\mu). $$

More details about the 5D differential calculus can be found
in Ref.~\cite{Agostini:2003dc} and references therein. For the purpose of
this article we quote the following commutation relation between one
forms $dx_a$ and \kM\ functions
 \be \Phi\,dx_b \mathcal{D}^b\Psi=dx^b
[\mathcal{D}_b(\Phi\Psi)-
(\mathcal{D}_b\Phi)\Psi].\label{diff_calc_compl}\ee They can be
easily proven by the iterative use of Eq.~(\ref{diff_comm_rel}).

\section{Noether's Theorem for Translations in \kM\ Spacetime}

The formulation of Noether's theorem in commutative spacetime is
recalled in Appendix~A. In this section we shall generalize
Noether's theorem for translations in \kM.

We consider the infinitesimal translation in \kM\ \be
\kx_\mu\to\kx'_\mu=\kx_\mu+dx_\mu,\ee where $dx_\mu$ are the
infinitesimal displacements. In order to ensure that the point
$\kx'_\mu$ still belongs to the \kM\ spacetime \bea
{}[\kx'_0,\kx'_j]=i\lambda \kx'_j,\;\;[\kx_j,\kx_k]=0\nn\eea the
translation parameters $dx_\mu$ must satisfy non-zero commutation
relations with the \kM\ generators $\kx_\mu$, namely
$$
[\kx_0,dx_j]+[dx_0,\kx_j]=i\lambda dx_j
$$
$$
[\kx_j,dx_k]+[dx_j,\kx_k]=0
$$
There are different choices for the commutation relations
$[\kx_\mu,dx_\nu]$ which fulfill the equations above. The various
choices select different choices of differential calculi in \kM\
(see, for example, Ref.~\cite{kappanoether1} where the infinitesimal
displacement is denoted by $\varepsilon_\mu$). In this paper we
chose the 5-dimensional differential calculus~(\ref{diff_comm_rel}).

As the spacetime changes also the field changes in order to preserve
the relativistic invariance. The total variation can be written as
the sum of two tems\footnote{we consider the approximation
$\delta\Phi(\kx')=\delta\Phi(\kx)$ at the first order in
$\varepsilon$.} \be
\delta_T\Phi(\kx)=\Phi'(\kx')-\Phi(\kx)\approx\delta\Phi(\kx)+dx^a
\mathcal{D}_a\Phi,\label{sum}\ee where $\delta\Phi(\kx)$ is the
contribution of the change of the form of the field and $dx^a
D_a\Phi$ is the contribution of the change of the argument $\kx$.

The first step for generalizing the results of Appendix~A to \kM\
consists in introducing an action for free scalar particles in \kM.
In order to do this we need to introduce a Lagrangian density and
define an integration map.

A good candidate for a covariant generalization of the Lagrangian
for scalar particles is
 \be
 \mathcal{L}(\kx)=\frac{1}{2}\left(
\tilde{\mathcal{D}}_a\Phi(\kx)\cdot\mathcal{D}^a\Phi(\kx)-m^2\Phi^2(\kx)\right)\label{lagrangian}\ee
with $\tilde{\mathcal{D}}^a\equiv
\frac{i}{\lambda}\left(\sinh(\lambda
P_0)-\frac{\lambda^2}{2}e^{\lambda P_0}\vec{P}^2,\lambda
P^j,i(2\sinh^2(\frac{\lambda P_0}{2})-\frac{\lambda^2}{2} e^{\lambda
P_0}\vec{P}^2)\right)$. In the next section we will see that the
Lagrangian~(\ref{lagrangian}) gives rise to \kP\ covariant equation
of motion (EoM). The choice of a Lagrangian which produces
\kP-covariant EoM is fundamental to assure a \kP-covariant
formulation of Noether theorem.

Concerning the integration map, a natural \kP\ translation-invariant
choice is (see Refs.~\cite{Majidqreg}, \cite{AmelinoCamelia:1999pm}) \be
\int \Phi(\kx)=\int \de^4x\, \phi(x),\label{integration}\ee where the
\kM\ function $\Phi(\kx)$ is written in the time-to-the-right
ordering (\emph{i.e.} in terms of the time-to-the-right basis
$\hat{e}_k$, see Eq.~(\ref{Phi})) and the right side is the usual
integration of the underlying commutative function $\phi(x)$. This
prescription is such that the integral of a partial derivative of a
suitably decaying function $\phi$ vanishes.

The next four properties of the integration we have considered will
be of precious help.
\begin{itemize}
\item The integral of an element of the \kM\ basis is a standard
Dirac function \be \quad \int \hat{e}_{k}=\delta(k). \ee
\item The following cyclicity property holds
\be \int [\Phi\Psi+\Psi\Phi]=\int [(1+e^{3\lambda
P_0})\Phi]\Psi.\label{cyclicity}\ee
\item The following integrations by parts hold\footnote{One can
prove it by using the following rule of integration by parts
$$0=\int P_\mu (\Phi\Phi)=\int \Delta(P_\mu)\cdot(\Phi\otimes
\Phi)=\int (P_\mu\Phi)\Phi+\int (e^{-\lambda P_0\delta_{\mu
j}}\Phi)(P_\mu \Phi).$$} \be \int \tilde{\mathcal{D}}^a\Phi\cdot
\Psi=-\int \Phi\cdot \mathcal{D}^a\Psi, \label{integrbypart}\ee \be
\int \Phi\cdot \mathcal{D}^a\mathcal{D}_a\Psi=\int
\mathcal{D}^a\mathcal{D}_a\Phi\cdot \Psi,\label{integrbypart1}\ee
where $\tilde{\mathcal{D}}^a=\frac{i}{\lambda}\left(\sinh(\lambda
P_0)-\frac{\lambda^2}{2}e^{\lambda P_0}\vec{P}^2,\lambda
P^j,i(2\sinh^2(\frac{\lambda P_0}{2})-\frac{\lambda^2}{2} e^{\lambda
P_0}\vec{P}^2)\right)$.
\item The space integral of a divergence $\int \de^3\kx P_j\Psi(\kx)$ is
zero:  \be\int \de^3 P_j (\Psi(\kx))=\int d^4k \psi(k) \left(\int
\de^3 P_j e^{ik\kx}\right)e^{-ik_0\kx_0}=\int d^4k \psi(k)
k_j\delta^3(k)e^{-ik_0\kx_0}=0. \label{divergence}\ee
\end{itemize}

The invariance under translation can be easily
proven\footnote{Observe that the step $\int dx^a f(\kx)=dx^a\int
f(\kx)$ is well defined even though $dx^a$ does not commute with
$f(\kx)$: one can show indeed (see Appendix~C in
Ref.~\cite{Agostini:2003dc}) that $[dx^a,f(\kx)]=O(P_\mu)$ thus, for any
decaying function $f$, $\int [dx^a, f(\kx)]=0$ and the integration
map can be applied directly to the function $f(\kx)$.} by using
property~(i)  $$ \int [\Phi'(\kx)-\Phi(\kx)]=-dx^a\int \mathcal{D}_a
\Phi(\kx)=-dx^a\tilde{\phi}(p) \mathcal{D}_a(p) \delta(p)=0.$$

The action of the theory is obtained by integrating the Lagrangian
density~(\ref{lagrangian})
 $$
S[\Phi]=\int \mathcal{L}(\kx)=\frac{1}{2} \int\left(
\tilde{\mathcal{D}}_a\Phi(\kx)\cdot\mathcal{D}^a\Phi(\kx)-m^2\Phi^2(\kx)\right)$$
with the integral map defined in Eq.~(\ref{integration}) and the
operator $\tilde{\mathcal{D}}$ introduced in Eq.~(\ref{integrbypart}).

Next, we get the EoM for free scalar particles in \kM. EoM are used
to obtain the conserved currents for \emph{on-shell} particles.

\subsection{$\kappa$-Poincar\'{e}-covariant Equations of Motion in \kM}

In this section we show that the Lagrangian~(\ref{lagrangian}) gives
rise to $\kappa$-covariant equation of motion (EoM) in \kM. In this
way we intend the $kappa$-covariance of the Lagrangian.

Let us consider an arbitrary variation $\delta\Phi$. Hamilton's
principle states that \be \delta S[\Phi]=\frac{1}{2}\int
[(\tilde{\mathcal{D}}_a\Phi)\mathcal{D}^a\delta\Phi+(\tilde{\mathcal{D}}_a\delta\Phi)\mathcal{D}^a\Phi-m^2\Phi\cdot\delta\Phi-m^2\delta\Phi\cdot\Phi]=0.\ee
By using properties~(\ref{integrbypart}),~(\ref{integrbypart1})
and~(\ref{cyclicity}), we obtain the EoM: \bea 0&=&\int
\Phi[(\mathcal{D}_a \mathcal{D}^a+m^2) \delta\Phi]+\delta\Phi(\mathcal{D}_a\mathcal{D}^a\Phi+m^2)\nn\\
&=&\int(\mathcal{D}_a \mathcal{D}^a+m^2)\Phi\cdot\delta\Phi+\delta\Phi\cdot(\mathcal{D}_a\mathcal{D}^a+m^2)\Phi\nn\\
&=&\int(1+e^{3\lambda
P_0})(\mathcal{D}_a\mathcal{D}^a+m^2)\Phi\cdot\delta\Phi
\quad\to\quad (\mathcal{D}_a\mathcal{D}^a+m^2)\Phi_0=0, \nn\\
\eea which contains a covariant generalization of D'Alembert's
operator of order two in the generalized derivatives $\mathcal{D}_a$
\be \Box_\lambda=\mathcal{D}_a
\mathcal{D}^a=-(\frac{4}{\lambda^2}\sinh^2(\lambda
P_0/2)-\vec{P}^2e^{\lambda P_0}),\ee which coincides with the
Casimir operator~(\ref{casimir})
and turns out to be covariant (under the generators $T$ of \kP) $$
[T,\mathcal{D}_a \mathcal{D}^a]=0 .$$ Of course, in the commutative
limit, $\Box_\lambda$ reduces to the commutative D'Alembert operator
$$ \mathcal{D}_a \mathcal{D}^a\to
\partial_\mu\partial^\mu.$$
Thus the Lagrangian~(\ref{lagrangian}) gives rise to EoM invariant
under the action
of the \kP\ generators.

\subsection{Conserved charges for \kM\ translations}

In this section we obtain the conserved currents for translations in \kM. We
show here the main steps of the procedure. The details of the
calculation can be found in Appendix~B.

As in the commutative case, the symmetry condition of the action can
be formulated as \bea \delta S_\Omega[\Phi]=\frac{1}{2}\int_{\Omega}
\left[\mathcal{L}(\Phi'(\kx'))-\mathcal{L}(\Phi(\kx))\right]=0.\nn
\eea

According to the variation of $\Phi$, Eq.~(\ref{sum}), the variation of
the action has two contributions \bea \delta
S_\Omega[\Phi]=\frac{1}{2}\int_{\Omega}
\left[\delta\mathcal{L}+dx_a\mathcal{D}^a\mathcal{L}\right].\nn \eea

Considering the variation of the action on the EoM, and after some
calculations, we can put the variation in the form (see appendix for
details) $$ \delta S_\Omega[\Phi]=dx^b \int_\Omega D^\mu J_{\mu
b}.$$

The invariance with respect to spacetime translations gives rise to
the continuity equation  \be
 D^\mu(P) J_{\mu b}=0,\label{continuityeq}\ee
 where $D^\mu=$ is a function of momenta and $J_{\mu b}$ is a five-dimensional generalization of the energy-momentum
 tensor.

The nutshell \kM-particles\footnote{We call nutshell \kM-particles
the free scalar particles $\Phi_0$ which satisfy the EoM in \kM\
spacetime $(\Box_\lambda+m^2)\Phi_0=0$.} can be written as \be
\Phi_0=\int d^4k
\;\tilde{\phi}(k)\hat{e}_k\delta(C_\lambda(k)),\label{nutshell}\ee
where $C_\lambda(k)=\frac{4}{\lambda^2}\sinh^2(\frac{\lambda
k_0}{2})-e^{\lambda k_0}k^2-m^2$.

The explicit form of $J_{\mu b}$ for on-shell particle $\Phi_0$ is
obtained in Appendix~C and corresponds to \bea
J_{\mu\bt}&=&e^{-\lambda P_0}\mathcal{D}_\bt \Phi_0\cdot
D_\mu\Phi_0+e^{-\lambda P_0\delta_{\bt0}}\tilde{D}_\mu\Phi_0\cdot
\mathcal{D}_\bt \Phi_0-i\delta_{\bt0}\lambda \tilde{D}_j
\tilde{D}_\mu\Phi_0\cdot D_j\Phi_0+\nn\\
&&+ig_{\mu0}\lambda m^2 e^{-\lambda P_0}\mathcal{D}_\bt\Phi_0\cdot
e^{\lambda P_0}\Phi_0-g_{\mu \bt}(1+e^{-\lambda P_0})^{\delta_{\bt0}}\mathcal{L}_{\Phi_0}\nn\\
J_{\mu4}&=&e^{-\lambda P_0}\mathcal{D}_4\Phi_0\cdot
D_\mu\Phi_0+e^{-\lambda P_0}\tilde{D}_\mu\Phi_0\cdot
\mathcal{D}_4\Phi_0-\lambda\tilde{D}_\mu\tilde{D}^\nu\Phi_0\cdot
D_\nu\Phi_0+\nn\\
&&+ig_{\mu 0}\lambda m^2e^{-\lambda P_0}\mathcal{D}_4\Phi_0\cdot
e^{\lambda P_0}\Phi_0+g_{\mu0}\lambda\tilde{D}_0
\mathcal{L}_{\Phi_0}\nn\eea

By spacial integration of $D^\mu J_{\mu b}$ we obtain \be D^0\int
\de^3 J_{0b}=\int \de^3 D_j J_{jb}=0, \label{conteq}\ee where we
 used the property that the space integral of a divergence is zero, see Eq.~(\ref{divergence}).

Thus $J_{0b}$ are the conserved currents which we are looking for.
On the EoM, the components of the current $J_{0b}$ are (see the
detailed computation in Appendix~C) \bea
J_{00}&=&\tilde{D_0}\Phi\cdot D_0\Phi+\tilde{D}_j \Phi\cdot D_j\Phi+
m^2 e^{-\lambda P_0}\Phi\cdot e^{\lambda P_0}\Phi-iJ_{04}\nn\\
 J_{0j}&=&\tilde{D}_j \Phi\cdot
D_0\Phi+\tilde{D}_0\Phi\cdot D_j \Phi+i\lambda m^2
\tilde{D}_j\Phi\cdot e^{\lambda
P_0}\Phi\nn\\
J_{04}&=&\frac{\lambda m^2}{2}(\tilde{D}_0+i\lambda m^2)(e^{-\lambda
P_0}\Phi\cdot e^{\lambda
P_0}\Phi)-\lambda\tilde{D}_0\tilde{D}^\nu\Phi\cdot
D_\nu\Phi+\lambda\tilde{D}_0 \mathcal{L}. \nn\eea

Eq.~(\ref{conteq}) implies that the quantity $Q_a=\int d^3x J_{0a}$,
corresponding to the conserved charges, is $\kx_0$-independent. We
show the time-independence in Appendix~D, where an explicit
construction of the conserved charges is obtained. The explicit form
of the conserved charges is \be Q_b=-\frac{i}{2}\int
d^4p\;\tilde{\phi}(\dot{-}p)\,\tilde{\phi}(p)\,e^{2\lambda
p_0}\,\tilde{{\mathcal{D}}}_b(p)\,\mbox{sgn}\left(\frac{2}{\lambda^2}(1-e^{-\lambda
p_0})+m^2\right)\delta\left[C_\lambda(p)\right] \label{Q}\ee where
$\mathcal{D}_b(p)$ is defined as in the Eq.~(\ref{relations}) and
sgn$(y)=\frac{y}{|y|}$ is the sign function. This form shows the
time-independence of the charges $Q_b$.

In this way we have obtained \emph{five} conserved-charges on the
equations of motion. In the limit $m=0$ the fifth charge $Q_4$ is
zero for on-shell massless particles, and we have four-conserved
charges.

In the next section we show that the analysis in terms of plane
waves allows us to recognize a relation between conserved charges.
This relation can have the meaning of dispersion relation for
particles and it allows to identify the energy-momentum vector.

\subsection{Plane waves and energy-momentum conservation law}

In Appendix~E we have proved that the Fourier transform of a real
on-shell \emph{plane wave} in \kM\ is described by the function \be
\tilde{\phi}_p(k)=\delta^{(3)}(k-p)H\!\!\left[k_0-\ln(1+\frac{\lambda^2m^2}{2})\right]+\delta^{(3)}(k\dot{+}p)H\!\!\left[-k_0-\ln(1+\frac{\lambda^2m^2}{2})\right]\label{planew_p}\ee
and we have also proved that \be
\tilde{\phi}_p(\dot{-}k)=e^{-3\lambda
k_0}\tilde{\phi}_p(k)\label{planew_p-}.\ee

By substituting Eq.~(\ref{planew_p}) and Eq.~(\ref{planew_p-}) in
Eq.~(\ref{Q}), we can compute the five-conserved charges \bea
Q_b&=&-\frac{i}{2}\int
d^4k\;\tilde{\phi}(\dot{-}k)\,\tilde{\phi}(k)\,e^{-\lambda
k_0}\,\tilde{{\mathcal{D}}}_b(k_0,\vec{k})\,\mbox{sgn}\left(\frac{2}{\lambda^2}(1-e^{-\lambda
p_0})+m^2\right)\delta[C_\lambda(k)]\nn\\
&=&-i\int d^4k\;e^{-\lambda
k_0}\left[\delta^{(3)}(k-p)H(k_0-\ln(1+\frac{\lambda^2m^2}{2}))\right]^2\tilde{{\mathcal{D}}}_b(k_0,\vec{k})\cdot\nn\\
&&\quad\quad\cdot\mbox{sgn}\left(\frac{2}{\lambda^2}(1-e^{-\lambda
p_0})+m^2\right)\delta[C_\lambda(k)]\nn\\
&=&-iV\frac{\lambda}{2}\int dk_0\;e^{-\lambda
k_0}\,\tilde{{\mathcal{D}}}_b(k_0,\vec{p})\;\frac{\delta[k_0-w_+(p)]}{|1+\frac{\lambda^2m^2}{2}-e^{-\lambda k_0}|}\nn\\
&=&\frac{-i\frac{\lambda V}{2}\,e^{-\lambda
w_+(p)}}{|1+\frac{\lambda^2m^2}{2}-e^{-\lambda
w_+(p)}|}\tilde{{\mathcal{D}}}_b\left(w_+(p),\vec{p}\right).\nn\eea
Thus, the conserved charges are proportional to
$\tilde{\mathcal{D}}_b(w_+(p),\vec{p})$:
$Q_b\sim\tilde{\mathcal{D}}_b(w_+(p),\vec{p})$ and, because of the
following relation \bea
[\tilde{\mathcal{D}}_\mu\tilde{\mathcal{D}}^\mu](w_+(p),\vec{p})&=&[\tilde{D}_0(w_+(p),\vec{p})+i\frac{\lambda m^2}{2}]^2-\tilde{D}_j^2\nn\\
&=&\tilde{D}_\mu\tilde{D}^\mu+i\lambda
m^2\tilde{D}_0-\frac{\lambda^2 m^4}{4}=-e^{-\lambda
w_+(p)}m^2+i\lambda m^2\tilde{D}_0-\frac{\lambda^2
m^4}{4}\nn\\
&=&-e^{-\lambda w_+(p)}m^2- m^2(1-e^{-\lambda
w_+(p)})-\frac{\lambda^2 m^4}{4}=- (\frac{4}{\lambda^2
m^2}+1)\frac{\lambda^2m^2}{4}\nn\\
&=&-\left(\frac{4}{\lambda^2
m^2}+1\right)\tilde{\mathcal{D}}_4^2(w_+(p),\vec{p}),\nn\eea the
relation among the charges is\footnote{This result has been also
obtained, independently, by G. Amelino-Camelia \emph{et al.} in
Ref.~\cite{AmelinoCamelia:2007vj}.} \be Q_\mu
Q^\mu+(1+\frac{4}{\lambda^2m^2})Q_4^2=0.
\label{new_dispersion_relation}\ee

Notice that in the massless case ($m=0$) the relation~(\ref{new_dispersion_relation}) reads as $$ Q_\mu Q^\mu=0$$ Thus, if
we give $Q_\mu$ the meaning of the energy-momentum vector, the
relation above says that the dispersion relation for massless
Klein-Gordon particles in \kM\ spacetime is not-deformed and
coincides with its special-relativistic limit.

In the case of massive particles a fifth conserved charges ($Q_4$)
appears. If we define
$\hat{Q}_b=i^{\delta_{b4}}(1+\frac{4}{\lambda^2m^2})^{\frac{\delta_{b4}}{2}}Q_b$,
all the charges are real and we get the relation \be \hat{Q}_\mu
\hat{Q}^\mu=\hat{Q}_4^2 \ee which might be viewed as a
special-relativistic dispersion-relation with a generalized mass
term represented by $\hat{Q}_4$.

\section{Comparison with Previous Results}

To our knowledge, the first exploratory analysis of Noether's
theorem in noncommutative spacetime appeared in
ref.~\cite{kappanoether1}. In this paper the study of translation
symmetries has been applied to the example of \kM\ noncommutative
spacetime. According to some algebraic arguments, the symmetries of
\kM\ should be described in terms of a Plank-scale-deformation of
the Poincar\'{e} algebra, and a Planck-scale-deformation should
affect the particle dispersion relations. The paper
Ref.~\cite{kappanoether1} was aimed at establishing whether these formal
observations about the presence of nonclassical symmetries might
have been confirmed by a physical perspective based on Noether's
analysis.

By considering a much used (four dimensional) differential calculus
in \kM\ spacetime, the presence of some nonlinearity in the
energy-momentum relation for scalar fields did emerge in
Ref.~\cite{kappanoether1}.  For massless particles it has been obtained
the following dispersion relations \be
\frac{4}{\lambda^2}\sinh^2{\frac{\lambda E}{2}}-e^{\lambda
E}\sum_{i=1}^3P^2_i=0,\label{old}\ee where $E,P_i$ are the energy
and momenta of the physical particle.

However it is natural to wonder if this result might depend on the
choice of the differential calculus and to search for an alternative
formulation based on a different calculus.

The present paper has generalized the study of Ref.~\cite{kappanoether1}
by replacing the four dimensional differential calculus by the five
dimensional \kP\ covariant differential calculus. This replacement
should guarantee a \kP-covariant formulation of Noether's theorem.
As in Ref.~\cite{kappanoether1}, our study has been focused on
translations for a scalar field theory in \kM\ spacetime. Our
analysis has revealed that the relation~(\ref{old}) does depend on
the choice of the differential calculus. Indeed, using a
\kP-covariant differential calculus we have found classical
properties for the energy-momentum charges which, in the massless
case, turn out to satisfy the special-relativity relation
$$
E^2-\sum_{i=1}^3P_i=0.$$

\section{Conclusions}

In this paper we have constructed a \kP\ covariant formulation of
Noether's theorem for translations in \kM. In order to guarantee the
\kP\ covariance of the formulation we have based our analysis on the
five dimensional covariant differential calculus. We have obtain
exactly the special-relativistic dispersion relations for massless
free scalar particles, while for massive particles we have found a
dispersion relation similar to the special-relativistic one but with
a further (fifth) conserved charge which might have the meaning of a
generalized mass.

Our results does not agree with the \kM\ deformed dispersion
relation conjectured on the basis of some heuristic arguments and
widely used in literature. However, the fact that in this first
exploratory application of our description of symmetries we have
only considered a free scalar theory in \kM\ spacetime might be a
significant limitation.
One may think that the theory considered here does not have enough
structure to give proper physical significance to energy-momentum;
such hypothesis may deserve attention for future investigations. It
would be of interest to explore whether massless particles in \kP\
spacetime be affected by a Planck-scale-deformation of the
special-relativistic dispersion relations if one attempts to extend
our result to the case of more structured theories, such as
interacting theories or gauge theories.

We remarque that in this paper only translations are considered. In
principle, the same method we have used for translations can be
applied to the Lorentz transformations. Future studies may wish to
investigate Lorentz transformations in \kM\ and the construction of
conserved angular momentum (if it exist).

\section*{Acknowledgments}

I would like to thank G.~Amelino-Camelia for many ideas and
inspiring discussions.


\setcounter{section}{0}

\appendix{Noether Theorem in Commutative Spacetime \label{Aa}} Classical
Noether's theorem states that symmetry properties of the Lagrangian
(or Hamiltonian) imply the existence of conserved quantities (see
for example Ref.~\cite{Goldstein}).

In order to formulate Noether's theorem we need the equation of
motion (EoM). For the sake of simplicity, we consider the case of a
scalar field $\phi(x)$.

Let us introduce the Lagrangian density
$\mathcal{L}(\phi,\partial_\mu\phi,x_\mu)$ which is, in general, a
function of the fields $\phi$ as well of the field derivatives
$\partial_\mu\phi$, and in general, might well be an explicit
function of $x_\mu$.

We derive the EoM through Hamilton's principle by variation of the
action, \emph{i.e.} the integral of
$\mathcal{L}(\phi,\partial_\mu\phi,x_\mu)$, over a region in
four-space \be S=\int_\Omega \de^4x\,
\mathcal{L}(\phi,\partial_\mu\phi,x_\mu). \ee We consider an
arbitrary variation on $\phi$ and $\partial_\mu\phi$, which are
taken to be zero at the bounding surface $\Gamma(\Omega)$.
Hamilton's principle states that $$ \delta S=\int_\Omega \de^4x\,
\delta\mathcal{L}=0.$$

The variation $\delta\mathcal{L}$ of the Lagrangian can be written
as $$
\delta\mathcal{L}(\phi,\partial_\mu\phi,x_\mu)=\frac{\partial\mathcal{L}}{\partial\phi}
\delta\phi+\frac{\partial\mathcal{L}}{\partial\phi_{,\mu}}
\delta\partial_\mu\phi,$$ or equivalently, by using the linearity of
$\delta$ with respect to the derivative
$\delta\partial_\mu\phi=\partial_\mu\delta\phi$,
 \be
\delta\mathcal{L}(\phi,\partial_\mu\phi,x_\mu)=\frac{\partial\mathcal{L}}{\partial\phi}
\delta\phi+\frac{\partial\mathcal{L}}{\partial(\partial_\mu\phi)}
\partial_\mu\delta\phi.\label{L_variation}\ee
Thus \bea \delta S&=&\int_\Omega
\de^4x\,\left\{\frac{\partial\mathcal{L}}{\partial\phi}
\delta\phi+\frac{\partial\mathcal{L}}{\partial(\partial_\mu\phi)}
\partial_\mu\delta\phi\right\}\nn\\
&=&\int_\Omega
\de^4x\,\left\{\frac{\partial\mathcal{L}}{\partial\phi}
-\partial_\mu\frac{\partial\mathcal{L}}{\partial(\partial_\mu\phi)}
\right\}\delta\phi+\int_\Omega
\de^4x\,\partial_\mu\left\{\frac{\partial\mathcal{L}}{\partial(\partial_\mu\phi)}
\delta\phi\right\}\label{hamilton}\\
&=&\int_\Omega
\de^4x\,\left\{\frac{\partial\mathcal{L}}{\partial\phi}
-\partial_\mu\frac{\partial\mathcal{L}}{\partial(\partial_\mu\phi)}
\right\}\delta\phi=0,\nn\eea where the second integral in
Eq.~(\ref{hamilton}) can be transformed by a four-dimensional divergence
theorem into an integral over the surface $\Gamma(\Omega)$ bounding
the region $\Omega$ and then vanishes since
$\delta\phi|_{\Gamma(\Omega)}=0$.

By requiring that $\delta S=0$ for any arbitrary variation $\delta
\phi$, we get the EoM \be
\frac{\partial\mathcal{L}}{\partial\phi}=\frac{d}{dx_\mu}\frac{\partial\mathcal{L}}{\partial(\partial_\mu\phi)}
\label{EOM}\ee which is satisfied by on-shell particles.

We now recall the formulation of Noether's theorem for on-shell particles, which is
useful in order to extend the construction to the noncommutative
case.

Noether's theorem applies to continuous transformations, and here we
are dealing only with them. Symmetry under coordinate transformation
refers to the effects of an infinitesimal transformation of the form
$$ x_\mu\to x_\mu'=x_\mu+\delta x_\mu,$$ where the infinitesimal
change $\delta x_\mu$ may be a function of all the other coordinates
$x_\nu$. The effect of a transformation in the fields themselves may
be described by $$ \phi(x_\mu)\to
\phi'(x'_\mu)=\phi(x_\mu)+\delta_{T}\phi(x_\mu),$$ where the total
variation $\delta_T\phi(x)$ results from changes of both the form
and the argument of the function $\phi(x)$.

As a consequence of the transformations of both the coordinates and
fields, the Lagrangian appears, in general, as a different function
of both the spacetime coordinates and the fields $$
\mathcal{L}(\phi,\partial_\mu\phi,x_\mu)\to
\mathcal{L'}\left(\phi'(x'),\partial_\mu\phi'(x'),x'_\mu\right). $$

The symmetry or invariance condition of the Lagrangian, can be
generalized at the level of the action integral, so that the
invariance of the magnitude of the action integral under the
transformation leads to the existence of conserved quantities $$
\delta_T S[\phi]=\int_{\Omega'}\de^4x\,
\mathcal{L'}\left(\phi'(x),\partial_\mu\phi'(x),x\right)
-\int_{\Omega}\de^4x\,
\mathcal{L}\left(\phi,\partial_\mu\phi,x\right)=0,$$ where $\Omega$
is an arbitrary region in the 4-D spacetime, and $\Omega'$ its
transformation.

By means of some computations we get $$ 0=\int_{\Omega}\de^4x
\left\{ \mathcal{L}(\phi'(x))-\mathcal{L}(\phi(x))+\partial_\mu
[\mathcal{L}\delta x^\mu]\right\}.$$ The variation $\delta_T S$ has
two contributions: the integral of $\delta_\phi\mathcal{L}(\phi)$ is
the variation of the action due to the variation of the functional
form of the field\footnote{which affects also the derivatives of the
fields $\phi_{,\mu}$.}, while the integral of $\partial_\mu
[\mathcal{L}\delta x^\mu]$ comes from the variation in the
four-volume $\delta\Omega$ \be 0=\int_{\Omega}\de^4x \left\{
\delta_\phi\mathcal{L}(\phi)+\partial_\mu [\mathcal{L}\delta
x^\mu]\right\}\label{deltaS_comm}.\ee

Writing the variation $\delta_\phi\mathcal{L}(\phi)$ as in
Eq.~(\ref{L_variation}) and integrating by part, we get $$
0=\int_{\Omega}\de^4x \left\{
\left(\frac{\partial\mathcal{L}}{\partial\phi}-\partial_\mu\frac{\partial\mathcal{L}}{\partial(\partial_\mu\phi)}\right)
\delta\phi+\partial_\mu\left(\frac{\partial\mathcal{L}}{\partial(\partial_\mu\phi)}
\delta\phi+\mathcal{L}\delta x^\mu\right)\right\}.$$

The first integral is zero for on-shell particles because of EoM
Eq.~(\ref{EOM}), and the formulation of the invariance reads \be
0=\int_{\Omega}\de^4x \frac{d}{dx_\mu}\left\{
\frac{\partial\mathcal{L}}{\partial\phi_{,\mu}}
\delta\phi+\mathcal{L}\delta
x_\mu\right\},\quad\phi\;\mbox{on-shell}.\ee Since the previous
equality holds for any $\Omega$, we have the continuity equation for
the current $J_\mu=\frac{\partial\mathcal{L}}{\partial\phi_{,\mu}}
\delta\phi+\mathcal{L}\delta x_\mu$$$ \frac{d}{dx_\mu}\left\{
\frac{\partial\mathcal{L}}{\partial\phi_{,\mu}}
\delta\phi+\mathcal{L}\delta x_\mu\right\}=0.$$ The continuity
equation tells us that if we integrate this current over a
space-like slice, we get a conserved quantity called the Noether
charge.

As a consequence of Noether's theorem the invariance of physical
systems with respect to the 10-Poincar\'{e} transformations gives
the law of conservation of 10-quantities. Here we focus only on
translations. The invariance with respect to spacetime translations
gives the well known law of conservation of energy-momentum.

A finite translation $x_\mu\to x'_\mu=x_\mu+a_\mu$ induces the
following field transformation $$
\phi'(x)=\phi+\delta\phi=\phi(x)-a^\mu\partial_\mu\phi(x).$$ Noether's theorem
states then the existence of a energy-momentum tensor $T_{\mu\nu}$
which satisfies the continuity equation $$ 0=\partial^\mu
T_{\mu\nu}= \frac{d}{dx_\mu}\left\{
\frac{\partial\mathcal{L}}{\partial\phi_{,\mu}}\phi_{,\nu}-g_{\mu\nu}\mathcal{L}\right\}.
$$ This equation describes the conservation of \emph{four} energy-momentum
charges
 $$ Q_\mu=\int \de^3x T_{0\nu}=\int \de^3x \left\{
\frac{\partial\mathcal{L}}{\partial\phi_{,0}}\phi_{,\nu}-g_{0\nu}\mathcal{L}\right\}.
$$

It is easy to show that the energy-momentum tensor for a real, free
scalar field $\phi(x)$ described by a Lagrangian density $
\mathcal{L}=\frac{1}{2}\partial_\mu\phi\partial^\mu\phi$ is of the
form
\be T_{\mu\nu}=
\frac{\partial\mathcal{L}}{\partial\phi_{,\mu}}\phi_{,\nu}-g_{\mu\nu}\mathcal{L}=\partial_\mu\phi\partial_\nu
\phi-g_{\mu\nu}\mathcal{L}.\label{NT_classic}\ee In the following
section, we shall construct a formulation of Noether's theorem in \kM\ which
provides us a generalization of this result.

\appendix{Properties of the 5D Vector Fields on the Solutions of the EoM \label{Ab}}

We notice that the deformed derivatives take the following form on
the solutions of the EoM ($\Phi_0:\;(\Box_\lambda+m^2) \Phi_0=0$)
\bea \mathcal{D}_a\Phi_0&=&\left\{\frac{i}{\lambda}\left[e^{\lambda
P_0}-(1+\frac{\lambda^2}{2}m^2)\right],iP_je^{\lambda P_0},\frac{\lambda}{2}m^2\right\}\Phi_0,\nn\\
\tilde{\mathcal{D}}_a\Phi_0&=&\left\{-\frac{i}{\lambda}\left[e^{-\lambda
P_0}-(1+\frac{\lambda^2}{2}m^2)\right],iP_j,-\frac{\lambda}{2}m^2\right\}\Phi_0.\label{D}\eea

It is useful to introduce the following vector fields (which
correspond to Eq.~(\ref{D}) in the limit $m=0$) \bea
D_a&=&\left\{\frac{i}{\lambda}\left[e^{\lambda
P_0}-1\right],ie^{\lambda P_0}P_j,0\right\}\nn\\
\tilde{D}_a&=&\left\{\frac{i}{\lambda}e^{-\lambda
P_0}\left[e^{\lambda P_0}-1\right],iP_j,0\right\}=e^{-\lambda
P_0}D_a\label{Dm0}\eea

The following relations hold \bea
\mathcal{D}_a\Phi_0&=&\left[D_0-i\frac{\lambda}{2}m^2,D_j,\frac{\lambda}{2}m^2\right]\Phi_0\nn\\
\tilde{\mathcal{D}}_a\Phi_0&=&\left[\tilde{D}_0+i\frac{\lambda}{2}m^2,\tilde{D}_jF,-\frac{\lambda}{2}m^2\right]\Phi_0\label{relations}\eea

It is easy to find the coproduct
of $D_\mu$ and $\tilde{D}_\mu$ on $\Phi_0$
\be\Delta(D)_\mu=D_\mu\otimes e^{\lambda P_0}+1\otimes
D_\mu.\label{coproductD}\ee
\be\Delta(\tilde{D}_\mu)=\tilde{D}_\mu\otimes 1+e^{-\lambda
P_0}\otimes \tilde{D}_\mu.\label{coproductDtilde}\ee

Observe that the action for on-shell particles can be written as\bea
S[\Phi_0]&=&\frac{1}{2}\int[
\tilde{\mathcal{D}}^a\Phi_0\cdot\mathcal{D}_a\Phi_0-m^2\Phi_0^2]\nn\\
&=&\frac{1}{2}\int[
\tilde{D}^\mu\Phi_0D_\mu\Phi_0-m^2\Phi_0e^{\lambda P_0}\Phi_0]\eea

\appendix{$\kappa$-Deformed Conserved Currents \label{Ac}}

Let us consider the infinitesimal translation in \kM\ spacetime \be
\kx_\mu\to\kx'_\mu=\kx_\mu+dx_\mu\label{translation}\ee where $dx_\mu$
is the infinitesimal displacement.

The total variation $\delta_T\Phi(\kx)$ of the field $\Phi(\kx)$ under
Eq.~(\ref{translation}) can be written as \bea
\delta_T\Phi(\kx)&=&\Phi'(\kx')-\Phi(\kx)=[\Phi'(\kx')-\Phi(\kx')]+[\Phi(\kx')-\Phi(\kx)]\nn\\
&=&\delta\Phi(\kx')+\de\Phi\nn\\
&\approx&\delta\Phi(\kx)+dx^a \mathcal{D}_a\Phi\nn\eea where $\delta
\Phi(\kx')=\Phi'(\kx')-\Phi(\kx')\approx
\Phi'(\kx)-\Phi(\kx)=\delta\Phi(\kx)$ at the first order in
$\varepsilon$, and $\de\Phi=dx^a \mathcal{D}_a\Phi$ according to
Eq.~(\ref{5Ddiffcalc}) of the 5D differential calculus. Since for
scalar fields $\delta_T\Phi(\kx)=0$, we get the relation \be
\Phi'(\kx)=\Phi(\kx)-dx^a \mathcal{D}_a\Phi \label{expansion}\ee

Let us consider the \emph{formal} Action for 
free scalar particles in \kM\ spacetime at the finite 4-volume
$\Omega$ \be S[\Phi]=\frac{1}{2}\int_{\Omega}\left(
\tilde{\mathcal{D}}_a\Phi\cdot\mathcal{D}^a\Phi-m^2\Phi^2\right).
\ee

The translation-invariance of the action implies that the total
variation of the action under the transformation Eq.~(\ref{translation})
\bea \delta S_\Omega[\Phi]=\frac{1}{2}\int_{\Omega'}
(\tilde{\mathcal{D}}_a\Phi'\cdot\mathcal{D}^a\Phi'-m^2\Phi'^2)(\kx)-\frac{1}{2}\int_\Omega
(\tilde{\mathcal{D}}_a\Phi\cdot\mathcal{D}^a\Phi-m^2\Phi^2)(\kx)\nn
\eea be zero. Because of the translation-invariance of the integral,
we can write \bea \delta S_\Omega[\Phi]&=&\frac{1}{2}\int_{\Omega}
(\tilde{\mathcal{D}}_a\Phi'\cdot\mathcal{D}^a\Phi'-m^2\Phi'^2)(\kx-\epsilon)-\frac{1}{2}\int_\Omega
(\tilde{\mathcal{D}}_a\Phi\cdot\mathcal{D}^a\Phi-m^2\Phi^2)(\kx)\nn\\&=&\frac{1}{2}\int_{\Omega}
(\tilde{\mathcal{D}}_a\Phi'\cdot\mathcal{D}^a\Phi'-m^2\Phi'^2)(\kx)-\frac{1}{2}\int_\Omega
\de(\tilde{\mathcal{D}}_a\Phi'\cdot\mathcal{D}^a\Phi'-m^2\Phi'^2)(\kx)\nn\\
&-&\frac{1}{2}\int_\Omega
(\tilde{\mathcal{D}}_a\Phi\cdot\mathcal{D}^a\Phi-m^2\Phi^2)(\kx),\nn\eea
and we get \bea\delta S_\Omega[\Phi]
&=&-\frac{1}{2}\int_{\Omega}\left\{ \de(\tilde{\mathcal{D}}_a
\Phi\cdot \mathcal{D}^a\Phi-m^2\Phi^2)- 2dx^a \mathcal{D}_a
\mathcal{L}\right\}\nn.\eea

Let us consider the solution of EoM ($\Phi=\Phi_0$) in the previous
equation\bea \delta
S_\Omega[\Phi_0]&=&-\frac{dx_b}{2}\int_{\Omega}\left\{\mathcal{D}^b[\tilde{D}^a\Phi\cdot
D_a\Phi-m^2\Phi\cdot e^{\lambda P_0}\Phi]- 2\mathcal{D}^b
\mathcal{L}\right\}_{\Phi=\Phi_0}
.\label{deltaS}\eea where $a,b=0,1,2,3,4$.

Let us consider first the term in $dx_\bt$ ($b=\beta$) in
Eq.~(\ref{deltaS}).
 Using Eq.~(\ref{copr_D_mu}) we obtain \bea
\mathcal{D}^\bt[\tilde{D}^a\Phi\cdot D_a\Phi-m^2\Phi\cdot e^{\lambda
P_0}\Phi]=\mathcal{D}^\bt \tilde{D}^a\Phi\cdot e^{\lambda
P_0}D_a\Phi\!&\!-\!&\!m^2\mathcal{D}^\bt\Phi\cdot
e^{2\lambda P_0}\Phi\nn\\
+e^{-\lambda P_0\delta_{\bt0}}\tilde{D}^a\Phi\cdot \mathcal{D}^\bt
D_a\Phi\!&\!-\!&\!m^2e^{-\lambda P_0\delta_{\bt0}}\cdot
e^{\lambda P_0}\mathcal{D}^\bt \Phi\nn\\
-i\delta_{\bt0}\lambda \tilde{D}_j \tilde{D}^a\Phi\cdot
D_jD_a\Phi\!&\!+\!&\!i\delta_{\bt0}\lambda m^2\tilde{D}_j\Phi\cdot
e^{\lambda P_0}D_j\Phi.\nn\eea
Using the coproduct of $D^a$, Eq.~(\ref{coproductD}), and the equality
$D^aD_a\Phi_0=-m^2e^{\lambda P_0}\Phi_0$, we get \bea
\mathcal{D}^\bt \tilde{D}^a\Phi_0\cdot e^{\lambda
P_0}D_a\Phi_0&=&D_a[e^{-\lambda P_0}\mathcal{D}^\bt \Phi_0\cdot
D^a\Phi_0]+m^2e^{-\lambda P_0}\mathcal{D}^\bt \Phi_0\cdot e^{\lambda
P_0}\Phi_0\nn\\  e^{-\lambda P_0\delta_{\bt0}}\tilde{D}^a\Phi_0\cdot
\mathcal{D}^\bt D_a\Phi_0&=&D_a[e^{-\lambda
P_0\delta_{\bt0}}\tilde{D}^a\Phi_0\cdot \mathcal{D}^\bt
\Phi_0]+m^2e^{-\lambda P_0\delta_{\bt0}}\Phi_0\cdot e^{\lambda
P_0}\mathcal{D}^\bt \Phi_0\nn\\ \tilde{D}_j \tilde{D}^a\Phi_0\cdot
D_jD_a\Phi_0&=&D_a[\tilde{D}_j \tilde{D}^a\Phi_0\cdot
D_j\Phi_0]+m^2\tilde{D}_j\Phi_0\cdot e^{\lambda P_0}D_j\Phi_0\nn\eea
Thus, \bea \mathcal{D}^\bt[\tilde{D}^a\Phi_0\cdot
D_a\Phi_0-m^2\Phi_0\cdot e^{\lambda P_0}\Phi_0]=D_a\left[e^{-\lambda
P_0}\mathcal{D}^\bt \Phi_0\cdot D^a\Phi_0+e^{-\lambda
P_0\delta_{\bt0}}\tilde{D}^a\Phi_0\cdot \mathcal{D}^\bt
\Phi_0\right.\nn\\
-\left.i\delta_{\bt0}\lambda\tilde{D}_j \tilde{D}^a\Phi_0\cdot
D_j\Phi_0+ig_{a0}\lambda m^2e^{-\lambda
P_0}\mathcal{D}^\bt\Phi_0\cdot e^{\lambda P_0}\Phi_0\right].\nn\eea
Moreover, we can write $ \mathcal{D}^\bt=\frac{(1+e^{-\lambda
P_0})^{\delta_{\bt0}}}{2}D^\bt+i\delta_{\bt0}\frac{\lambda}{2}e^{\lambda
P_0}\vec{P}^2$ so that \be \mathcal{D}^\bt
\mathcal{L}[\Phi_0]=\frac{(1+e^{-\lambda
P_0})^{\delta_{\bt0}}}{2}D^\bt
\mathcal{L}[\Phi_0]+i\delta_{\bt0}\frac{\lambda}{2}e^{\lambda
P_0}\vec{P}^2\mathcal{L}[\Phi_0],\label{DL}\ee where we can ignore
the last term since it does not contribute to the conserved
quantities which are defined up to divergence terms, as in the
commutative case.

Using the same procedure for the term in $dx_4$ ($b=4$) in
Eq.~(\ref{deltaS}), we obtain \bea
\mathcal{D}^4[\tilde{D}^a\Phi_0\cdot D_a\Phi_0-m^2\Phi_0\cdot
e^{\lambda P_0}\Phi_0]=D_a\left[e^{-\lambda
P_0}\mathcal{D}^4\Phi_0\cdot D^a\Phi_0+e^{-\lambda
P_0}\tilde{D}^a\Phi_0\cdot
\mathcal{D}^4\Phi_0\right.\nn\\
-\left.\lambda\tilde{D}^\nu \tilde{D}^a\Phi_0\cdot
D_\nu\Phi_0+ig_{a0}\lambda m^2e^{-\lambda
P_0}\mathcal{D}^4\Phi_0\cdot e^{\lambda
P_0}\Phi_0\right]_{\Phi_0}.\nn\eea (let us remember that
$\mathcal{D}^4$ acts as a constant over $\Phi_0$:
$\mathcal{D}^4\Phi_0=\frac{\lambda}{2}m^2\Phi_0$).
Moreover, we can write $$\mathcal{D}^4
\mathcal{L}[\Phi_0]=-\frac{\lambda}{2}\tilde{D}_0D_0
\mathcal{L}[\Phi_0]+\frac{\lambda}{2}e^{\lambda
P_0}\vec{P}^2\mathcal{L}[\Phi_0],$$ where we can ignore the last
term as in Eq.~(\ref{DL}).

The variation of the action~(\ref{deltaS}) can be finally written as
\bea \delta S&=&-\frac{dx_\bt}{2} \int_\Omega D^\mu\left[e^{-\lambda
P_0}\mathcal{D}^\bt \Phi\cdot D_\mu\Phi+e^{-\lambda
P_0\delta_{\bt0}}\tilde{D}_\mu\Phi\cdot \mathcal{D}^\bt
\Phi-i\delta_{\bt0}\lambda \tilde{D}_j \tilde{D}_\mu\Phi\cdot
D_j\Phi\right.\nn\\
&&\quad\quad\quad+\left.ig_{\mu0}\lambda m^2 e^{-\lambda
P_0}\mathcal{D}^\bt\Phi\cdot e^{\lambda P_0}\Phi-g_{\mu \bt}(1+e^{-\lambda P_0})^{\delta_{\bt0}}\mathcal{L}\right]_{\Phi=\Phi_0}\nn\\
&&-\frac{dx_4}{2}\int_\Omega D^\mu\left[e^{-\lambda
P_0}\mathcal{D}^4\Phi\cdot D_\mu\Phi+e^{-\lambda
P_0}\tilde{D}_\mu\Phi\cdot
\mathcal{D}^4\Phi-\lambda\tilde{D}_\mu\tilde{D}^\nu\Phi\cdot
D_\nu\Phi\right.\nn\\
&&\quad\quad\quad\left.+ig_{\mu 0}\lambda m^2e^{-\lambda
P_0}\mathcal{D}^4\Phi\cdot e^{\lambda
P_0}\Phi+g_{\mu0}\lambda\tilde{D}_0
\mathcal{L}\right]_{\Phi=\Phi_0}\nn\eea

The continuity equation is $D^\mu J_{\mu b}=0$ where\bea
J_{\mu\bt}&=&e^{-\lambda P_0}\mathcal{D}_\bt \Phi_0\cdot
D_\mu\Phi_0+e^{-\lambda P_0\delta_{\bt0}}\tilde{D}_\mu\Phi_0\cdot
\mathcal{D}_\bt \Phi_0-i\delta_{\bt0}\lambda \tilde{D}_j
\tilde{D}_\mu\Phi_0\cdot D_j\Phi_0+\nn\\
&&+ig_{\mu0}\lambda m^2 e^{-\lambda P_0}\mathcal{D}_\bt\Phi_0\cdot
e^{\lambda P_0}\Phi_0-g_{\mu \bt}(1+e^{-\lambda P_0})^{\delta_{\bt0}}\mathcal{L}_{\Phi_0}\nn\\
J_{\mu4}&=&e^{-\lambda P_0}\mathcal{D}_4\Phi_0\cdot
D_\mu\Phi_0+e^{-\lambda P_0}\tilde{D}_\mu\Phi_0\cdot
\mathcal{D}_4\Phi_0-\lambda\tilde{D}_\mu\tilde{D}^\nu\Phi_0\cdot
D_\nu\Phi_0+\nn\\
&&+ig_{\mu 0}\lambda m^2e^{-\lambda P_0}\mathcal{D}_4\Phi_0\cdot
e^{\lambda P_0}\Phi_0+g_{\mu0}\lambda\tilde{D}_0
\mathcal{L}_{\Phi_0}\nn\eea
and the conserved currents are \bea J_{00}&=&e^{-\lambda
P_0}\mathcal{D}_0 \Phi_0\cdot D_0\Phi_0+e^{-\lambda
P_0}\tilde{D}_0\Phi_0\cdot \mathcal{D}_0 \Phi_0-i\lambda \tilde{D}_j
\tilde{D}_0\Phi_0\cdot D_j\Phi_0\nn\\
&&+i\lambda m^2 e^{-\lambda P_0}\mathcal{D}_0\Phi_0\cdot e^{\lambda
P_0}\Phi_0-(1+e^{-\lambda
P_0})\mathcal{L}_{\Phi_0}\nn\\
J_{0j}&=&\tilde{D}_j \Phi_0\cdot D_0\Phi_0+\tilde{D}_0\Phi_0\cdot
D_j \Phi_0+i\lambda m^2 \tilde{D}_j\Phi_0\cdot e^{\lambda
P_0}\Phi_0\nn\\
J_{04}&=&e^{-\lambda P_0}\mathcal{D}_4\Phi_0\cdot
D_0\Phi_0+e^{-\lambda P_0}\tilde{D}_0\Phi_0\cdot
\mathcal{D}_4\Phi_0-\lambda\tilde{D}_0\tilde{D}^\nu\Phi_0\cdot
D_\nu\Phi_0\nn\\
&&+i\lambda m^2e^{-\lambda P_0}\mathcal{D}_4\Phi_0\cdot e^{\lambda
P_0}\Phi_0+i(1-e^{-\lambda P_0})\mathcal{L}_{\Phi_0} \nn\eea
After some calculations they read as \bea
J_{00}&=&\tilde{D_0}\Phi_0\cdot D_0\Phi_0+\tilde{D}_j \Phi_0\cdot
D_j\Phi_0+
m^2 e^{-\lambda P_0}\Phi_0\cdot e^{\lambda P_0}\Phi_0-iJ_{04}\nn\\
 J_{0j}&=&\tilde{D}_j \Phi_0\cdot
D_0\Phi+\tilde{D}_0\Phi_0\cdot D_j \Phi_0+i\lambda m^2
\tilde{D}_j\Phi_0\cdot e^{\lambda
P_0}\Phi_0\nn\\
J_{04}&=&\frac{\lambda m^2}{2}(\tilde{D}_0+i\lambda m^2)(e^{-\lambda
P_0}\Phi_0\cdot e^{\lambda
P_0}\Phi_0)-\lambda\tilde{D}_0\tilde{D}^\nu\Phi_0\cdot
D_\nu\Phi_0+\lambda\tilde{D}_0 \mathcal{L}_{\Phi_0} \nn\eea

To simplify the calculation of the conserved charges $Q_b=\int
\de^3xJ_{0b}$ we introduce the following linear combinations of
$J_{0b}$: \bea K_{00}&=&J_{00}+iJ_{04}=\tilde{D}_0\Phi_0\cdot
D_0\Phi_0+\tilde{D}_j\Phi_0\cdot D_j\Phi_0+ m^2 e^{-\lambda
P_0}\Phi_0\cdot
e^{\lambda P_0}\Phi_0\nn\\
K_{0j}&=&J_{0j}=\tilde{D}_j \Phi_0\cdot
D_0\Phi_0+\tilde{D}_0\Phi_0\cdot D_j \Phi_0+i\lambda m^2
\tilde{D}_j\Phi_0\cdot e^{\lambda
P_0}\Phi_0\nn\\
K_{04}&=&J_{04}=\frac{\lambda m^2}{2}(\tilde{D}_0+i\lambda
m^2)(e^{-\lambda P_0}\Phi_0\cdot e^{\lambda
P_0}\Phi_0)-\lambda\tilde{D}_0\tilde{D}^\nu\Phi_0\cdot
D_\nu\Phi_0+\lambda\tilde{D}_0 \mathcal{L}_{\Phi_0}
\label{newcurrents}\eea and in the next section we compute the
conserved quantities $\tilde{Q}_b=\int \de^3x K_{0b}$. We will then
obtain $Q_b$ by linear combinations of $\tilde{Q}_b$ $$
Q_b=\tilde{Q}_b-i\delta_{b0}\tilde{Q}_4$$

\appendix{Construction of the Conserved Charges \label{Ad}}
By using the expression~(\ref{nutshell}) for the field $\Phi(\kx)$,
we can write the conserved quantities $\tilde{Q}_b$ associated with
the currents $K_{0b}$ \bea \tilde{Q}_b(\kx_0)=\int
K_{0b}(\kx)=\frac{1}{2}\int \de^3x\int d^4k d^4p\;
\;\tilde{\phi}(k)K_{0b}(k_\mu,p_\mu)\tilde{\phi}(p)\delta[C_\lambda(k)]\delta[C_\lambda(p)]\hat{e}_k\hat{e}_p\nn
\eea where $K_{0b}(k_\mu,p_\mu)$ represent the Fourier-transforms of
$K_{0b}(\kx)$ (\ref{newcurrents}).

The conserved quantities associated with the currents $K_{0b}$ are
\bea \tilde{Q}_b(\kx_0)=\int K_{0b}(\kx)=\frac{1}{2}\int \de^3x\int
d^4k d^4p\;
\;\tilde{\phi}(k)K_{0b}(k_\mu,p_\mu)\tilde{\phi}(p)\delta[C_\lambda(k)]\delta[C_\lambda(p)]\hat{e}_k\hat{e}_p\nn
\eea where \bea K_{00}(k_\mu,p_\mu)&=&\tilde{D}_0(k)
D_0(p)+\tilde{D}_j(k) D_j(p)+ m^2 e^{-\lambda k_0}e^{\lambda
p_0}\nn\\
K_{0j}(k_\mu,p_\mu)&=&\tilde{D}_j(k)D_0(p)+\tilde{D}_0(k)
D_j(p)+i\lambda m^2 \tilde{D}_j(k)e^{\lambda p_0}\nn\\
K_{04}(k_\mu,p_\mu)&=&\frac{\lambda m^2}{2}e^{\lambda
(p_0-k_0)}(\tilde{D}_0(k_0+p_0)+i\lambda
m^2)-\lambda\tilde{D}_0\tilde{D}^\nu\Phi\cdot
D_\nu\Phi\nn\\
&&+\frac{\lambda}{2}\tilde{D}_0(k_0+p_0)(\tilde{D}^\nu(k)D_\nu(p)-m^2e^{-\lambda
p_0}).\eea
Integrating in $\de^3x$ we obtain \bea
\tilde{Q}_b(\kx_0)=\frac{1}{2}\int d^4k
d^4p\;\;\tilde{\phi}(k)K_{0b}(k_\mu,p_\mu)\tilde{\phi}(p)\delta[C_\lambda(k)]\delta[C_\lambda(p)]\delta^{(3)}(k+e^{-\lambda
k_0} p)e^{i(k_0+p_0)\kx_0}\nn\eea and computing the $\delta^{(3)}k$
we get \bea \tilde{Q}_b(\kx_0)&=&\frac{1}{2}\int
dk_0d^4p\;\tilde{\phi}(k_0,-e^{-\lambda
k_0}p)K_{0b}(k_0,-e^{-\lambda
k_0}p,p_0,p)\tilde{\phi}(p)e^{i(k_0+p_0)\kx_0}\cdot\nn\\
&&\delta\left(\frac{4}{\lambda^2}\sinh^2(\frac{\lambda
k_0}{2})-e^{-\lambda
k_0}\vec{p}^2-m^2\right)\delta\left(\frac{4}{\lambda^2}\sinh^2(\frac{\lambda
p_0}{2})-e^{\lambda p_0}\vec{p}^2-m^2\right)\nn\eea
By considering the product of the two delta functions, it is easy to
see that only two solutions are possible $k_0^{(1)}=-p_0$ and
$k_0^{(2)}=\lambda^{-1}\ln(2+\lambda^2m^2-e^{-\lambda p_0})$.
However, we can prove that the second solution does not contribute
to the integral, in fact \bea K_{00}(k_0^{(2)},-e^{-\lambda
k_0^{(2)}}p,p_0,p)
&=&e^{-\lambda k_0}[-\frac{2}{\lambda^2}(\cosh(\lambda
p_0)-1)+e^{\lambda p_0}p_j^2+m^2]\nn\\
&=&-e^{-\lambda k_0}C_\lambda(p)=0\nn\\
K_{0j}(k_0^{(2)},-e^{-\lambda k_0^{(2)}}p,p_0,p)&=&0\nn\\
K_{04}(k_0^{(2)},-e^{-\lambda k_0^{(2)}}p,p_0,p)&=&0\nn \eea Thus,
for $k_0=k_0^{(2)}$, the integral is zero and we are left with the
solution $k_0=k_0^{(1)}=-p_0$. In this way the term dependent on
$\kx_0$ disappears and $\tilde{Q}_b$ turns out to be
time-independent. Thus, $\tilde{Q}_b$ take the form \bea
\tilde{Q}_b&=&\frac{1}{2}\int
dk_0d^4p\;\tilde{\phi}(\dot{-}p)K_{0b}(\dot{-}p_\mu,p_\mu)\tilde{\phi}(p)\delta\left(\frac{4}{\lambda^2}\sinh^2(\frac{\lambda
k_0}{2})-e^{\lambda
k_0}\vec{p}^2-m^2\right)\cdot\nn\\
&&\cdot\delta\left(\frac{4}{\lambda^2}\sinh^2(\frac{\lambda
p_0}{2})-e^{\lambda p_0}\vec{p}^2-m^2\right) \nn\eea and after some
calculations they become \bea \tilde{Q}_b&=&\frac{1}{2}\int
dk_0d^4p\;\tilde{\phi}(\dot{-}p)K_{0b}(\dot{-}p_\mu,p_\mu)\tilde{\phi}(p)\frac{\delta(k_0+p_0)}{\lambda|\frac{4}{\lambda^2}e^{-\frac{\lambda
p_0}{2}}\sinh(\frac{\lambda
p_0}{2})+m^2|}\cdot\delta[C_\lambda(p)]\nn\eea
One can verify that for $k_0=-p_0$ the functions
$K_{0b}(\dot{-}p_\mu,p_\mu)$ are\bea
K_{0\mu}(\dot{-}p_\mu,p_\mu)&=&-i\lambda e^{\lambda p_0}D_\mu(p)
(\frac{4}{\lambda^2}e^{-\frac{\lambda
p_0}{2}}\sinh(\frac{\lambda p_0}{2})+ m^2) \nn\\
K_{04}(\dot{-}p_\mu,p_\mu)&=&i\frac{\lambda^2m^2}{2}e^{2\lambda
p_0}(\frac{4}{\lambda^2}e^{-\frac{\lambda
p_0}{2}}\sinh(\frac{\lambda p_0}{2})+ m^2)\nn\eea
 Substituting this expression in $\tilde{Q}_b$ we
get \bea \tilde{Q}_\mu&=&-\frac{i}{2}\int
d^4p\;\tilde{\phi}(\dot{-}p)\,\tilde{\phi}(p) \,e^{\lambda
p_0}\,D_\mu(p) \,\mbox{sgn}\!\left(\frac{2}{\lambda^2}(1-e^{-\lambda
p_0})+m^2\right)\delta[C_\lambda(p)]\nn\\
\tilde{Q}_4&=&\frac{i}{2}\int
d^4p\;\tilde{\phi}(\dot{-}p)\,\tilde{\phi}(p) \,e^{2\lambda
p_0}\,\frac{\lambda m^2}{2}
\,\mbox{sgn}\!\left(\frac{2}{\lambda^2}(1-e^{-\lambda
p_0})+m^2\right)\delta[C_\lambda(p)]\nn\eea where
$\dot{-}p_\mu=(-p_0,-e^{\lambda p_0}p)$ is the antipode of $p_\mu$
and sgn$(y)=y/|y|$ is the sign function.

The five conserved charges $Q_b$ ($b=0,1,2,3,4.$) can be obtained by
linear combinations of $\tilde{Q}_b$: \be
Q_b=\tilde{Q}_b-i\delta_{b0}\tilde{Q}_4=-\frac{i}{2}\int
d^4p\;\tilde{\phi}(\dot{-}p)\tilde{\phi}(p)e^{2\lambda
p_0}\tilde{{\mathcal{D}}}_b(p)\mbox{sgn}(\frac{2}{\lambda^2}(1-e^{-\lambda
p_0})+m^2)\delta[C_\lambda(p)] \ee where
$\tilde{{\mathcal{D}}}_b(p)$ is defined as in the
Eq.~(\ref{relations}).

\section{$\kappa$-Minkowski Plane Waves\label{Ae}}

Let us consider the function $\tilde{\phi}(k)$ in Eq.~(\ref{nutshell})
given by \be
\tilde{\phi}_p(k)=N_p\delta^{(3)}(p-k)\,H\!\!\left[k_0-\ln(1+\frac{\lambda^2m^2}{2})\right],\label{pwave}\ee
where $H(x)$ is the Heaviside function (equal to 0 if $x\leq 0$ and
equal to 1 if $x>0$), and $N_p$ is a suitable normalization which
can be chosen in different ways as in the commutative case. To be
simple, let us set $N_p=1$.

In this appendix we want to show that the field $\Phi_0^{(p)}(\kx)$
\be \Phi_0^{(p)}(\kx)=\int \de^{4}k
\,\tilde{\phi_p}(k)\,e^{-ik\kx}e^{ik_0\kx_0}\,\delta[C_\lambda(p)]
\label{field}\ee corresponds to a on-shell \emph{plane wave} in \kM.
By substituting the function~(\ref{pwave}) in Eq.~(\ref{field}) we
obtain\bea \Phi_0^{(p)}(\kx)=\int d^4k
\delta^{(3)}(p-k)H\!\!\left[k_0-\ln(1+\frac{\lambda^2m^2}{2})\right]\delta\left(\frac{4}{\lambda^2}\sinh^2(\frac{\lambda
k_0}{2})-e^{\lambda k_0}k^2-m^2\right)e^{-ik\kx}e^{ik_0\kx_0}\nn\\
=\frac{\lambda}{2}\int dk_0
H\!\!\left[k_0-\ln(1+\frac{\lambda^2m^2}{2})\right]\left(\frac{\delta(k_0-w_+(p))}{|1+\frac{\lambda^2m^2}{2}-e^{-\lambda
w_+}|}+\frac{\delta(k_0-w_-(p))}{|1+\frac{\lambda^2m^2}{2}-e^{-\lambda
w_-}|}\right)e^{-ip\kx}e^{ik_0\kx_0} \nn\eea where
$w_\pm(p)=\lambda^{-1}\ln\left(\frac{1+\frac{\lambda^2m^2}{2}\pm\sqrt{\frac{\lambda^4m^4}{4}+\lambda^2(\vec{p}^2+m^2)}}{1-\lambda^2\vec{p}^2}\right)$.
Since $w_-(p)<\ln\left(1+\frac{\lambda^2m^2}{2}\right)$ then $$
\Phi_0^{(p)}(\kx)=\frac{\lambda}{2}\int dk_0
\frac{\delta(k_0-w_+(p))}{|1+\frac{\lambda^2m^2}{2}-e^{-\lambda
w_+(p)}|}e^{-ip\kx}e^{ik_0\kx_0}=\frac{\lambda}{2}\frac{
e^{-ip\kx}e^{iw_+(p)\kx_0}}{|1+\frac{\lambda^2m^2}{2}-e^{-\lambda
w_+(p)}|}$$ and we see that the field $\Phi_0^{(p)}$ corresponds to
a on-shell \emph{plane wave} in \kM.

The real version of the field $\Phi_0^{(p)}(\kx)$ is \bea
\Phi_0^{(p,real)}(\kx)&=&\frac{\lambda}{2}\frac{(e^{-ip\kx}e^{iw_+\kx_0}+h.c.)}{|1+\frac{\lambda^2m^2}{2}-e^{-\lambda
w_+(p)}|}=\frac{\lambda}{2}\frac{(e^{-ip\kx}e^{iw_+\kx_0}+e^{ipe^{\lambda
w_+}\kx}e^{-iw_+\kx_0})}{|1+\frac{\lambda^2m^2}{2}-e^{-\lambda
w_+(p)}|}\nn \eea where we have used the property~(\ref{conj})
$\hat{e}(k))^\dag=\hat{e}(\dot{-}k)$.

It is easy to see that the Fourier transform corresponding to the
real plane wave $\Phi_0^{(p,real)}(\kx)$ is
$$
\tilde{\phi}_p(k)=\delta^{(3)}(k-p)H\!\!\left[k_0-\ln(1+\frac{\lambda^2m^2}{2})\right]+\delta^{(3)}(k\dot{+}p)H\!\!\left[-k_0-\ln(1+\frac{\lambda^2m^2}{2})\right].$$
Under the transformation $k_\mu \to \dot{-}k_\mu$ we get \bea
\tilde{\phi}_p(\dot{-}k)
=e^{-3\lambda
k_0}\left[\delta^{(3)}(k\dot{+}p)H(-k_0-\ln(1+\frac{\lambda^2m^2}{2}))+\delta^{(3)}(k-p)H(k_0-\ln(1+\frac{\lambda^2m^2}{2}))\right].\nn\eea


\end{document}